\documentclass[12pt]{article}
\pdfoutput=1

\usepackage[bulletsep]{collref}

\usepackage[utf8]{inputenc}
\usepackage[T1]{fontenc}
\usepackage{amsmath,amssymb, amsfonts}
\usepackage{icomma}
\usepackage[utf8]{inputenc}
\usepackage[T1]{fontenc}
\usepackage{color}
\usepackage{graphicx}
\usepackage{titlesec}
\usepackage{lmodern}
\usepackage{units}
\usepackage{marvosym}
\usepackage{MnSymbol}
\usepackage[usenames,dvipsnames,svgnames,table]{xcolor}
\usepackage[small]{subfigure}

\newcommand{\beq}{\begin{equation}}
\newcommand{\eeq}{\end{equation}}

\newcommand{\sign}{\text{sign}}

\begin{document}

\renewcommand{\thefootnote}{\fnsymbol{footnote}}
\setcounter{footnote}{0}

\begin{flushright}\footnotesize
\texttt{NORDITA-2014-70} \\
\texttt{UUITP-06/14}
\vspace{0.6cm}
\end{flushright}

\begin{center}
{\Large\textbf{\mathversion{bold} 
Quantum Phase Transitions\\
 in Mass-Deformed ABJM Matrix Model
}
\par}

\vspace{0.8cm}

 \textrm{Louise~Anderson$^{1}$ and
Konstantin~Zarembo$^{2,3}$\footnote{Also at ITEP, Moscow, Russia}}
\vspace{4mm}

\textit{${}^1$Department of Fundamental Physics\\
Chalmers University of Technology\\
S-412 96 G\"oteborg, Sweden}\\
\textit{${}^2$Nordita, KTH Royal Institute of Technology and Stockholm University,
Roslagstullsbacken 23, SE-106 91 Stockholm, Sweden}\\
\textit{${}^3$Department of Physics and Astronomy, Uppsala University\\
SE-751 08 Uppsala, Sweden}\\
\vspace{0.2cm}
\texttt{louise.anderson@chalmers.se, zarembo@nordita.org}

\vspace{1em}
\begin{abstract}
When mass-deformed ABJM theory is considered on $S^3$, the partition function of the theory localises, and is given by a matrix model. At large $N$, we solve this model in the decompactification limit, where the radius of the three-sphere is taken to infinity. In this limit, the theory exhibits a rich phase structure with an infinite number of third-order quantum phase transitions,  accumulating at strong coupling. 
\end{abstract}

\end{center}

\newpage

\tableofcontents

\section{Introduction}

In supersymmetric gauge theories, the path integral can sometimes be evaluated exactly \cite{Witten:1988ze}. An interesting example of particular importance for our work is the partition function on the sphere for field theories with extended supersymmetry \cite{Pestun:2007rz}.
The method of localisation allows for observables with sufficient amount of supersymmetry to be written in terms of matrix integrals, which are immensely simpler than the original  functional integral expressions and yet carry a lot of information of the vacuum structure and the non-perturbative dynamics of  the underlying field theories. 

In the large $N$  (i.e. multicolour or planar) limit, the localisation matrix integrals can be analyzed by standard tools of random matrix theory \cite{Brezin:1977sv} and in some cases even solved exactly at any coupling. These results have many applications, including providing insights into the strong-coupling behaviour of field theories with holographic duals. The ability to compute quantities exactly on the gauge theory side then allows for direct comparisons with gravity- and string theory calculations for quantities that non-trivially depend on the coupling constant.

Herein, we focus on the mass-deformed ABJM model, a three-dimensional Chern-Simons theory with two gauge groups and matter in the bi-fundamental representation.  In the massless case, the ABJM theory enjoys an $\mathcal{N}=6$ superconformal symmetry and is dual to type IIA string theory on $AdS_4\times CP^3$ \cite{Aharony:2008ug}. Its partition functions on $S^3$ localises to a matrix model \cite{Kapustin:2009kz,Kapustin:2010xq}, exactly solvable at large $N$ \cite{Drukker:2010nc}, which makes possible a very detailed comparison between field theory calculations and geometric analyses in string theory \cite{Drukker:2010nc,Herzog:2010hf,Martelli:2011qj,Cheon:2011vi,Jafferis:2011zi,Drukker:2011zy,Fuji:2011km, Gulotta:2011aa, Alday:2012au, Assel:2012cp, Bhattacharyya:2012ye, Freedman:2013oja, Farquet:2013cwa, Hatsuda:2013oxa, Marmiroli:2013nza, Lewkowycz:2013laa, Bianchi:2014laa, Gromov:2014eha, Farquet:2014kma, Honda:2014npa, Dabholkar:2014wpa}. 

It is known that mass deformation away from the conformal point may lead to rather dramatic effects, especially in the decompactification limit when the radius of the sphere is taken to infinity. For instance,  the Chern-Simons theory coupled to massive fundamental matter then undergoes quantum weak/strong coupling phase transitions at some critical values of the 't~Hooft coupling \cite{Barranco:2014tla}, quite similar to the phase transitions found in the four-dimensional $\mathcal{N}=2$ Super-QCD \cite{Russo:2013kea}. 

Drawing further on the analogy with four-dimensional theories, one may expect that the theory with bi-fundamental matter will have a much richer phase structure. Such a theory in four dimensions, usually referred to as $\mathcal{N}=2^*$ super-Yang-Mills (SYM), undergoes an infinite number of phase transitions which accumulate at strong coupling \cite{Russo:2013qaa},  where the holographic duality is supposed to operate. What these phase transitions correspond to on the string theory side remains an open problem, partly because a complete analytic solution across the whole phase diagram is still missing. 

As we shall see, the mass-deformed ABJM theory displays a very similar behavior, undergoing infinitely many phase transitions as the coupling grows from zero to infinity. Moreover, the matrix model of  mass-deformed ABJM becomes exactly solvable  in the decompactification limit, allowing us to completely map the entire phase diagram of this model. 

The paper is organized as follows: in section \ref{sec:ContRank} we introduce the matrix model for mass-deformed ABJM theory, and discuss its decompactification limit. The saddle-point equations are then solved for two different analytic continuations of the original model in sec.~\ref{sec:exact-s} and sec.~\ref{sec:ContLevel}. The results are discussed in sec.~\ref{sec:discussion}.

\section{Massive ABJM \label{sec:ContRank}} 

The ABJM model is an $\mathcal{N}=6$ superconformal Chern-Simons theory  with the gauge group $U_{k}(N) \times U_{-k}(N)$  and matter in the bi-fundamental representation, where $k$ as usual denotes the Chern-Simons level. 
The path integral of this theory on  $S^3$ was shown to localise on constant field configurations in \cite{Kapustin:2009kz,Kapustin:2010xq,Jafferis:2010un,Hama:2010av}, but this result actually does not require as high supersymmetry as $\mathcal{N}=6$, nor relies on conformal invariance, and thus holds true in a much wider class of models, conformal or not. Localisation opens an avenue to study the large $N$ limit of these models  by methods of random matrix theory, a great simplification compared to direct summation of planar diagrams. The main effort, largely motivated by the AdS/CFT duality, has been directed towards conformal models (see \cite{Marino:2011nm} for a review), while much less is known about massive theories. In three dimensions, the Chern-Simons theory with fundamental matter remains the only case studied so far \cite{Barranco:2014tla}.  Herein, we concentrate on the mass deformation of the ABJM model  obtained by giving equal masses to all bi-fundamental fields.

We shall actually consider a small generalisation of  ABJM theory, in which the two gauge groups are allowed to have different ranks: $U_{k}(N_1) \times U_{-k}(N_2)$. This generalisation proved useful in the study of the massless theory \cite{Drukker:2010nc},  as it allows for varying the two 't~Hooft couplings independently. The localised path integral in this case becomes equivalent to the partition function of pure Chern-Simons theory on the lens space $L(2,1)$ \cite{Marino:2002fk,Aganagic:2002wv}. The two matrix models are related through changing the sign of $N_2$, making possible to exploit the large-$N$ solution of the lense-space matrix model \cite{Halmagyi:2003ze} in the ABJM context.

The localisation locus of ABJM theory consists of spatially homogeneous auxiliary fields, $\sigma $ and $\tilde{\sigma }$, from the vector multiplets of the two gauge groups, which can be brought to  diagonal form by a gauge transformation:
\begin{equation}
 \sigma =\mathop{\mathrm{diag}}\left(\mu _1,\ldots ,\mu _{N_1}\right),
 \qquad 
 \tilde{\sigma }=\mathop{\mathrm{diag}}\left(\nu _1,\ldots ,\nu _{N_2}\right).
\end{equation}
The partition function on $S^3$ then takes the form of an eigenvalue integral  \cite{Kapustin:2009kz,Kapustin:2010xq} (using the normalization convention of \cite{Drukker:2010nc}):
 \begin{align}
 \label{eq:matrix_model_scaled}
 Z_{ABJM}(\zeta, m, k) =& \frac{1}{N_1!\,N_2!} \int \prod_{i=1}^{N_1}\frac{ d\mu_i}{2 \pi}\,\,\prod_{a=1}^{N_2} \frac{d\nu_a}{2 \pi}\,\, 
 \\ \nonumber & 
 \times 
 \frac{\prod\limits_{i\neq j} \sinh^2\frac{\mu_i -\mu_j}{2} 
 \prod\limits_{a\neq b} \sinh^2\frac{\nu_a -\nu_b}{2}}
 {\prod\limits_{i\, a} \cosh\frac{\mu_i-\nu_a+ m}{2} \cosh\frac{\mu_i-\nu_a-m}{2}}\,\,
 \,{\rm e}\,^{i\zeta\left(
 \sum\limits_{i}^{}\mu _i+\sum\limits_{a}^{}\nu _a
 \right) + 
 \frac{ik   }{4 \pi} \left(
 \sum\limits_i \mu_i^2-\sum\limits_{a}^{}\nu_a^2
 \right)
 }
 \end{align}
 The mass-deformation is here represented by $m$, the Fayet-Illiopoulos parameter $\zeta$ is included for completeness and is set to zero in the rest of the paper.  

We will be interested in the ABJM model with $N\rightarrow \infty $. In the standard 't~Hooft limit, the Chern-Simons level should also be sent to infinity such that the  't~Hooft coupling is held fixed:
\begin{equation}
 \lambda =\frac{2\pi N}{k}\,.
\end{equation}
As mentioned above, for solving the model in the massless case it was actually easier to start with a more general model where the ranks of the gauge groups are analytically continued to pure imaginary values of the 't~Hooft couplings. We thus introduce
\begin{equation}
 \lambda _{1}=\frac{2\pi iN_{1}}{k}\,,\qquad 
  \lambda _{2}=-\frac{2\pi iN_{2}}{k}\,,
\end{equation}
and assume that $\lambda _{1,2}$ are real. The physical value of the 't~Hooft coupling in the original ABJM model is obtained by analytic continuation $\lambda _1\rightarrow \,{\rm e}\,^{i\varphi }\lambda $, $\lambda _2\rightarrow \,{\rm e}\,^{-i\varphi }\lambda $ with $\varphi $ going from $0$ to $\pi /2$.  The analytic continuation to complex $\lambda _{1,2}$ is however not at all straightforward  in the massive theory, which will be discussed more towards the end of the paper.

In the large $N$ limit, the saddle-point approximation for the eigenvalue integral (\ref{eq:matrix_model_scaled}) becomes exact.
The saddle-point equations for the $\mu_i$'s and $\nu_a$'s take the form:
\begin{align}
\label{eq:saddlepoints_partial_bac}
 \mu_i =
 &
  \frac{\lambda_1}{N_1} \sum_{ j \neq i} \coth\frac{\mu_i -\mu_j}{2} 
+  \frac{\lambda_2}{2 N_2}    \sum_{a}\Big(\tanh\frac{\mu_i-\nu_a+ m}{2} +\tanh\frac{\mu_i-\nu_a-m}{2} \Big)  \\ \nonumber 
\nu_a =
 &
   \frac{\lambda_2}{N_1}  \sum_{ b\neq a} \coth\frac{\nu_a -\nu_b}{2}  
+  \frac{\lambda_1}{2 N_2}   \sum_{i}\Big(\tanh\frac{\nu_a-\mu_i+ m}{2} +\tanh\frac{\nu_a-\mu_i-m}{2}  \Big) 
.\end{align}

The analytic continuation in $\lambda $, however, is not unique. We may as well start with the equations
\begin{align}
\label{eq:saddlepoints_partial_bac_acont}
 \frac{\mu_i}{\alpha _1} =
 &
  \frac{1}{N} \sum_{ j \neq i} \coth\frac{\mu_i -\mu_j}{2} 
 - \frac{1}{2 N}    \sum_{a}\Big(\tanh\frac{\mu_i-\nu_a+ m}{2} +\tanh\frac{\mu_i-\nu_a-m}{2} \Big)  \\ \nonumber 
\frac{\nu_a}{\alpha _2} =
 &
   \frac{1}{N}  \sum_{ b\neq a} \coth\frac{\nu_a -\nu_b}{2}  
-  \frac{1}{2 N}   \sum_{i}\Big(\tanh\frac{\nu_a-\mu_i+ m}{2} +\tanh\frac{\nu_a-\mu_i-m}{2}  \Big) 
,\end{align}
where we already assume that the two groups have equal rank. The original ABJM model is then obtained by analytically continuing in $\alpha _{1,2}$: $\alpha  _1\rightarrow \,{\rm e}\,^{i\varphi }\lambda $, $\alpha _2\rightarrow \,{\rm e}\,^{-i\varphi }\lambda $. These equations can be regarded as the saddle-point equations for the matrix model
 \begin{align}
 \label{eq:matrix_model_scaled_acont}
 Z= \frac{1}{N!^2} \int \prod_{i,a=1}^{N}\frac{ d\mu_i\, d\nu_a}{(2 \pi)^2}\,\, 
 \frac{\prod\limits_{i\neq j} \sinh^2\frac{\mu_i -\mu_j}{2} 
 \prod\limits_{a\neq b} \sinh^2\frac{\nu_a -\nu_b}{2}}
 {\prod\limits_{i\, a} \cosh\frac{\mu_i-\nu_a+ m}{2} \cosh\frac{\mu_i-\nu_a-m}{2}}\,
 \,{\rm e}\,^{- 
 N  \left(
 \frac{1}{2\alpha  _1}\sum\limits_i \mu_i^2+\frac{1}{2\alpha  _2}\sum\limits_{a}^{}\nu_a^2
 \right)
 }.
 \end{align}

Loosely speaking, the first case, that leads to \eqref{eq:saddlepoints_partial_bac}, can be regarded as analytic continuation in the rank of the gauge group, while the second case can be interpreted as analytic continuation in  the Chern-Simons level. In this paper, we shall investigate both these cases. In neither one, the saddle-point equations are, to our knowledge, possible to solve analytically, but may be investigated through numerical methods. Solutions may be found when both $\lambda_1, \lambda_2 \in \mathbb{R}$ (or $\alpha _1, \alpha  _2 \in \mathbb{R}$ ), whereas the solutions quickly become unstable  when the 't~Hooft couplings acquire complex phases. A more thorough discussion on these features shall be given in the conclusions, and we shall for the moment not dwell on the analytic continuation any further.

\subsection{The decompactification limit \label{sec:DecompLimit}}

The localisation formulae above are written in units where the radius of the sphere is set to one. The dependence on $R$ can be reinstated by rescaling all dimensionful variables:
\begin{equation}\label{eq:limit}
 m\rightarrow mR, \qquad \mu _i\rightarrow \mu _iR,\qquad \nu _a\rightarrow \nu _aR.
\end{equation}
An obviously interesting question is what happens when the radius of the sphere goes to infinity.   Apart from simplifying the saddle-point equations, this limit brings in new qualitative features. Massive theories  in four dimensions, when decompactified, appear to undergo phase transitions at some critical values of the 't~Hooft coupling \cite{Russo:2013qaa} (see \cite{Russo:2013sba} for a review). Phase transitions of this type was also observed in the Chern-Simons theory with fundamental matter \cite{Barranco:2014tla}, in close analogy to $\mathcal{N}=2$ QCD in four dimensions \cite{Russo:2013kea,Russo:2013sba}. We expect that the phase structure of the mass-deformed ABJM resembles that of the four-dimensional $\mathcal{N}=2^*$ theory, (since they both are theories with matter in the bifundamental representation), which is substantially more complicated, with an infinite number of phase transitions accumulating at strong coupling \cite{Russo:2013qaa,Russo:2013kea}. We will find that the saddle-point equations  simplify in the decompactification limit to the extent that they may be studied by analytical means herein.

A quick inspection of the saddle-point equations (\ref{eq:saddlepoints_partial_bac}) or  (\ref{eq:saddlepoints_partial_bac_acont}) demonstrates that  simply rescaling (\ref{eq:limit}) and then taking $R\rightarrow \infty $  is not self-consistent. To circumvent this problem, it was suggested in \cite{Barranco:2014tla} that the 't~Hooft coupling should also be rescaled with $R$. Unlike (\ref{eq:limit}), this rescaling does not follow from dimensional analysis, and thus introduces a new scale to the problem, defined as 
\begin{equation}
 t=\frac{\lambda }{R}\,.
\end{equation}
This scale is kept fixed as $R\rightarrow \infty $. 

Since we have introduced partial 't Hooft couplings in the analytically continued saddle-point equations, there will be two different dimension-one parameters, $t_1$ and $t_2$ (corresponding to $\lambda_1, \lambda_2$ respectively). 
In terms of these, the saddle-point equations \eqref{eq:saddlepoints_partial_bac} will in the decompactification limit take the form\footnote{Here we assume that $t_1$ and $t_2$ are real and so are the eigenvalues. For a more general case of complex couplings and eigenvalues, $\mathop{\mathrm{sign}}z$ should be understood as $\mathop{\mathrm{sign}}\mathop{\mathrm{Re}}z$. The appearance of the real part indicates that analyticity is actually lost in the decompactification limit.}:
\begin{align}
\label{eq:saddlepoints_decomp}
&&\mu_i =
  \frac{ t_1}{N} \sum_{ j \neq i} \sign(\mu_i -\mu_j) 
+  \frac{ t_2}{2 N}    \sum_{a}\Big(\sign(\mu_i-\nu_a+ m)
 +\sign(\mu_i-\nu_a-m) \Big)  \\ \nonumber 
&&\nu_a =
   \frac{t_2}{N}  \sum_{ b\neq a} \sign(\nu_a-\nu_b) 
+ \frac{ t_1}{2 N}   \sum_{i}\Big(\sign(\nu_a-\mu_i+ m)
 +\sign(\nu_a-\mu_i-m) \Big) 
,\end{align}
where we used that $\tanh Rz$ and $\coth Rz$ becomes  step functions as $R\rightarrow \infty $. This step-function approximation obviously leads to drastic simplifications, and is actually familiar from the study of the conformal ABJM model \cite{Suyama:2009pd,Herzog:2010hf,Suyama:2011yz,Giasemidis:2013oea}, where this approximation corresponds to the extreme strong-coupling limit.

We should stress that the massless ABJM model with physical couplings $t_2=-t_1\gg 1$ is  a very special case since it results in  perfect cancellations on the right-hand side in equation \eqref{eq:saddlepoints_decomp}, which largely determine the structure of the solution at strong coupling. 
 In the cases we consider however, the mass shifts in the argument of the $\sign$-functions, together with the lack of any imposed conditions on the couplings, will result in a  situation where such cancellations do not occur. The solution for the cases considered herein, as a result, is thus expected to behave quite differently in comparison to the massless ABJM case.

After solving the saddle-point equations for real $t_1$, $t_2$,  one may try to rotate these real couplings into the complex plane by giving them phases of equal magnitude but opposite signs, and then letting these approach $\pm \frac{\pi}{2}$, corresponding to the case of real Chern-Simons level and ranks of the gauge groups. However, the hyperbolic functions in equation (\ref{eq:saddlepoints_partial_bac}) develop poles at purely imaginary argument, which is an indication that the analytical continuation back to the physical values of the couplings might not be as straight-forward as hoped. 

\section{Exact solution}\label{sec:exact-s}

\subsection{Some  simple examples \label{sec:ExamplesPlusPlus}}

Perhaps the most interesting feature of the model under consideration is the appearance of phase transitions at finite values of the 't~Hooft coupling. We first illustrate this phenomenon in the simplest examples, and then solve the model in full generality for the case of equal couplings ($t_1=t_2$).

It is convenient to introduce the eigenvalue densities for $\mu$ and $\nu$:
  \begin{align}
  \rho_{\mu   }(\mu )  = \frac{1}{N_1} \sum_i \delta(\mu  -\mu _i) ,
  \qquad 
  \rho _{\nu }(\nu )=\frac{1}{N_2}\sum_{a}^{}\delta (\nu -\nu _a)
 ,\end{align}
which as usual satisfies the normalisation condition 
 \beq
 \label{eq:norm_cond}
 \int \rho_\mu (\mu ) d \mu  =1=\int \rho_\nu (\nu ) d \nu.
 \eeq
The saddle-point equations \eqref{eq:saddlepoints_decomp} then become integral equations for the densities:
\begin{align}
\label{eq:saddlepoints_cont}
\mu =
 &
   t_1 \int_{\mathcal{C}_\mu} d\mu'\,\rho _\mu (\mu ') \sign(\mu -\mu') 
\nonumber \\
&+  \frac{ t_2}{2}    \int_{\mathcal{C}_\nu} d \nu\,\rho _\nu (\nu ) \Big(\sign(\mu-\nu+ m) 
 +\sign(\mu-\nu-m)   \Big)  \\ \nonumber 
\nu=
 &
 t_2  \int_{\mathcal{C}_\nu} d \nu' \,\rho (\nu ')\sign (\nu -\nu')  
 \nonumber \\
&+ \frac{ t_1}{2 }  \int_{\mathcal{C}_\mu} d\mu \,\rho (\mu ) \Big(\sign( \nu-\mu+ m)
 +\sign(\nu-\mu-m)  \Big) 
,\end{align}
where $\mathcal{C}_\mu$ and $\mathcal{C}_\nu$ denote the intervals on which $\rho_\mu$ and $\rho_\nu$ are supported respectively: $\mathcal{C}_\mu = [-A,A]$ and similarly $\mathcal{C}_\nu = [-B,B]$.

By differentiating the equations in \eqref{eq:saddlepoints_cont} with respect to $\mu$ and $\nu$ respectively, these are simplified considerably:
 \begin{align}
\label{eq:eig_densities_real}
 \rho_\mu(\mu)    =   &      
\frac{1}{2 t_1  }
-  \frac{t_2}{2 t_1 }      \Big( \rho_\nu(\mu+ m)  +  \rho_\nu(\mu- m) \Big) \\
\label{eq:eig_densities_real1}
 \rho_\nu(\nu)  =   &       
\frac{1}{ 2 t_2  }
-\frac{t_1}{2  t_2 }     \Big(  \rho_\mu(\nu+m) +  \rho_\mu(\nu-m)   \Big) 
,\end{align}
where the eigenvalue densities are taken to vanish outside their regions of support.  Since the two equations are symmetric under the exchange of $t_1$ and $t_2$ and $\mu $ with $\nu $, we may without loss of generality assume $A\geq B$. 

The eigenvalue densities thus satisfy a set of coupled finite-difference equations.
As one can see by trial and error, the only sensical solution is a constant or a  piecewise constant density. The precise appearance will differ depending on the value of $m$ in relation to the interval lengths, and will undergo abrupt changes when certain resonance conditions are fulfilled. Let us consider some simple examples:

\paragraph{
Decoupled solution: $A+B < m$.} 
\begin{figure}
\centering
\includegraphics[width=0.7\textwidth]{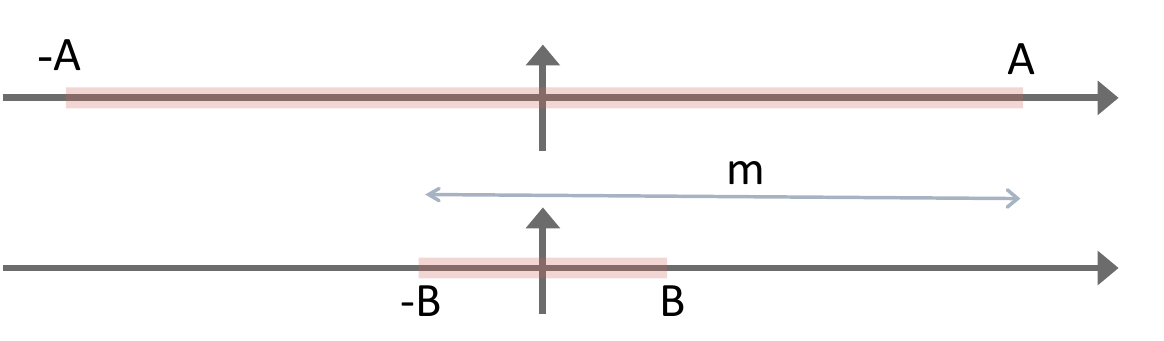}
\caption{{\small The case when $A+B < m$ and the interval on which the two eigenvalues are supported. The upper graph represents $\rho_{\mu}$ and the lower $\rho_{\nu}$.}}
\label{fig:m_geq_AB}
\end{figure}

In this case, $\mu \pm m$ and $\nu \pm m$ lie outside the intervals $\mathcal{C}_\mu $, $\mathcal{C}_\nu $ whenever $\mu \in [-A,A]$, $\nu \in [-B,B]$. The saddle-point equations of \eqref{eq:eig_densities_real}  and \eqref{eq:eig_densities_real1}  then decouple and the solution is simply
 \begin{align}
 \rho_\mu(\mu)    =         
\frac{1}{ 2  t_1 }\, \qquad 
 \rho_\nu(\nu)  =         
\frac{1}{ 2  t_2  }\,.
\end{align}
This behavior is easy to understand graphically (fig.~\ref{fig:m_geq_AB}): the interaction offsets the intervals by $\pm m$ and for $A+B<m$ there is no overlap between the resulting intervals.
The endpoints $A$ and $B$ can be found from the normalisation condition of \eqref{eq:norm_cond}:
 \begin{align}
A  =         
  t_1,\qquad  
B=         
 t_2 
.\end{align}
The solutions thus holds for
\beq 
t_1+t_2 < m.
\eeq
It may be noted that this solution corresponds to two decoupled Chern-Simons theories at large imaginary 't~Hooft coupling. 

\paragraph{Solution with resonances:  $A< m < A+B$.}
\begin{figure}
\centering
\includegraphics[width=0.7\textwidth]{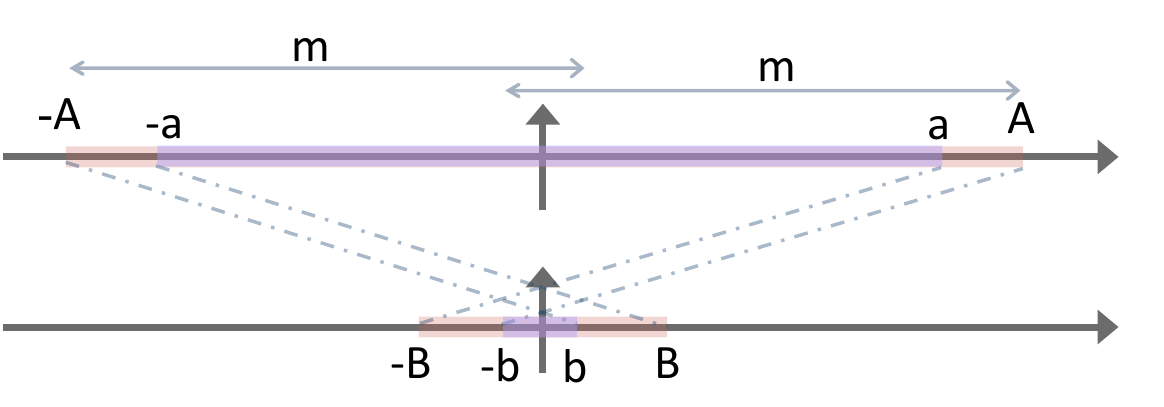}
\caption{ {\small The case when $A< m < A+B$. $a$ and $b$ are the resonance points.}}
\label{fig:m_geq_A}
\end{figure}

As the mass decreases, the offset intervals move closer to one another and at some point start to overlap. In the overlap region, the interaction terms in (\ref{eq:eig_densities_real}), (\ref{eq:eig_densities_real1}) are no longer zero. As a consequence, the eigenvalue distributions develop two patches with unequal densities. The density experiences a jump at the resonances, a point distance $m$ away from an endpoint of the other interval.

 In this situation, each one of the intervals will be divided into three regions, but due to reflection symmetry around the origin, there are in practise only two distinctly different regions of each interval where the values of the eigenvalue densities differ (fig.~\ref{fig:m_geq_A}). 
Consider, for instance the interval $[-A,A]$. This is split into two regions:  $|\mu| \in [0,a]$ and $|\mu| \in [a,A]$, where $a$ is given by the condition $a=m-B$. Similarly, the interval $[-B,B]$ is split into three regions $[-B,b]$, $[-b,b]$ and $[b,B]$ with $b=m-A$

The saddle-point equations of \eqref{eq:eig_densities_real} and \eqref{eq:eig_densities_real1} determine the eigenvalue densities in the different regions as:
\begin{align}
 \rho_\mu(\mu)    =   &      
 \begin{cases}
\frac{1}{ 2   t_1 }  \hspace*{2cm} |\mu| \in [0, a]
 \\
\frac{1}{ 3  t_1 } \hspace*{2cm} |\mu| \in [a, A]
\end{cases}
\\
 \rho_\nu(\nu)  =   &       
 \begin{cases}
\frac{1}{ 2  t_2  }  \hspace*{2cm} |\nu| \in [0, b]
 \\
\frac{1}{ 3  t_2 } \hspace*{2cm} |\nu| \in [b, B]
\end{cases}
\end{align}
Using the normalisation requirements for the eigenvalue densities, we find:
\begin{align}
A=&   t_2+2t_1-m\\ \nonumber
B=&    t_1+2t_2-m
.\end{align}
The assumption $A \geq B$ is equivalent to $t_1 \geq t_2$, and so the condition for  $A < m < A+B$ is fulfilled as soon as 
\begin{align}
t_1+t_2 > m,\qquad t_2+2t_1 < 2m.
\end{align}
 
\paragraph{Solution with secondary resonances: $\max(B, 2A/3) < m < A$. }

\begin{figure}
\centering
\includegraphics[width=0.7\textwidth]{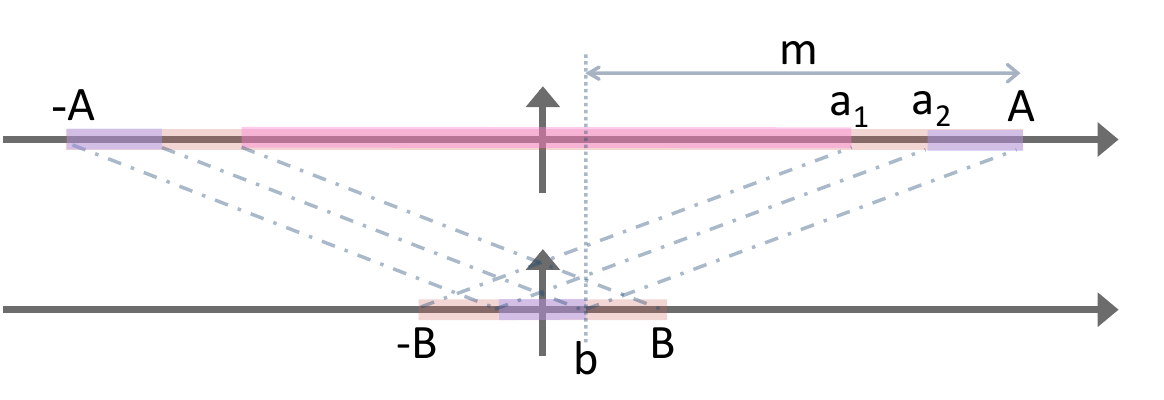}
\caption{{\small The case when $B< m < A$. $a_2$ is a secondary resonance.}}
\label{fig:m_geq_B}
\end{figure}

This situation is slightly more complicated than the ones previously considered. 
The solution of $\rho_\nu(\nu)$ for $|\nu|$ in the interval $[0,B]$ will have a discontinuity at the point $b=A-m$. In the region $|\nu| \in [b,B]$, (denoted by red in figure \ref{fig:m_geq_B}), one of the resonance points $\nu \pm m$ will lie inside $[-A,A]$, (more precisely in $[-a_2,-a_1] \cup [a_1,a_2]$, where $a_1=m-B$ and $a_2=2m-A$ ). For $|\nu| \in [0,b]$, (in figure \ref{fig:m_geq_B} denoted by purple), both resonance points $\nu \pm m$ will lie inside $[-A,A]$, with one in $[-A,-a_2]$ and the other in $[a_2,A]$. 

In a similar fashion, the interval $[0,A]$ is divided into three parts: the region $[0,a_1]$  where both resonance points will lie outside $[-B,B]$ (denoted by pink in figure \ref{fig:m_geq_B}), the region $[a_1,a_2]$  where one resonance point, $\mu -m$, will lie in $[-B,-b]$ (denoted by red), and finally, there is the region $\mu \in [a_2,A]$ (denoted by purple), where again one resonance point, $\mu-m$, will lie inside the support of $\rho_\nu$, but this time rather in $[-b,b]$ than in the previous region where it would lie in the outermost regions of $[-B,B]$.

Again, by considering the equations \eqref{eq:eig_densities_real} and \eqref{eq:eig_densities_real1} in the different regimes, one finds:
\begin{align}
 \rho_\mu(\mu)    =   &      
 \begin{cases}
\frac{1}{ 2   t_1 }  \hspace*{2cm} |\mu| \in [0, a_1]
 \\
\frac{1}{ 3  t_1 } \hspace*{2cm} |\mu| \in [a_1, a_2]
 \\
\frac{1}{ 2 t_1 } \hspace*{2cm} |\mu| \in [a_2, A]
\end{cases}
\\
 \rho_\nu(\nu)  =   &       
 \begin{cases}
0  \hspace*{2.6cm} |\nu| \in [0, b]
 \\
\frac{1}{ 3  t_2 } \hspace*{2cm} |\nu| \in [b, B]
\end{cases}
\end{align}
Once more, equation  \eqref{eq:norm_cond} allows us to relate the interval endpoints to the couplings:
\begin{align}
\label{eq:case3_ints}
A=  t_1+\frac{t_2}{2}\hspace*{5mm}
B=t_1+2t_2-m.
\end{align}
This solution exists for $t_2+2t_1>2m$ and $ t_1+2t_2<2 m$ (when $2A/3<B<m$) or $2t_1+t_2 < 3m$ (when $B<2A/3<m$).

\paragraph{Solution with a pair of secondary resonances: {\small $ B> 2A/3 ~~ \&  ~~  2A/3<m < A$.}}
\begin{figure}
\centering
\includegraphics[width=0.7\textwidth]{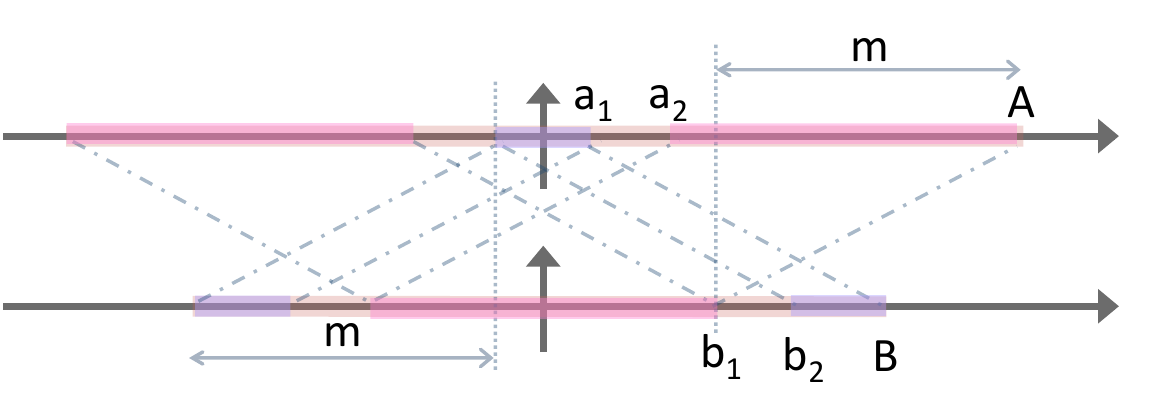}
\caption{{\small The case when $B> 2A/3 \hspace*{2mm} \&  \hspace*{2mm}   2A/3<m < A$ when both intervals carry primary and secondary resonances.}}
\label{fig:B_geq_05A_m_leq_A}
\end{figure}

This situation is similar to the previous case, but herein, both intervals $[0,A]$ and $[0,B]$ will be divided into three pieces, as illustrated in figure \ref{fig:B_geq_05A_m_leq_A}. The eigenvalue densities in the different regions may be found in a straight-forward manner, and are given by:
\begin{align}
\rho_\mu(\mu) =& 
\begin{cases}
0 \hspace*{17mm}  |\mu| \in [0,a_1] \\ 
 \frac{1}{3   t_1} \hspace*{11.5mm}   |\mu| \in [a_1,a_2] \\ 
  \frac{1}{2   t_1} \hspace*{11.5mm}    |\mu| \in [a_2,A] 
\end{cases}
\end{align}
\begin{align}
\rho_\nu(\nu) =& 
\begin{cases}
0 \hspace*{14mm}  |\mu| \in [0,b_1] \\ 
 \frac{1}{3   t_1} \hspace*{11.5mm}    |\nu| \in [b_1,b_2] \\
  \frac{1}{2   t_1} \hspace*{11.5mm}    |\nu| \in [b_2,B] 
  \end{cases}
.\end{align}
The normalisation conditions gives:
\begin{align}
A= t_1+\frac{t_2}{2}
\hspace*{1cm} &\hspace*{1cm}  B=t_2+\frac{t_1}{2}\,.
\end{align}
The upper limit on the mass for the existence of this solution is naturally given by the lower limit of the previous cases, whereas the lower limit may be expressed as
\beq
 t_1+\frac{t_2}{2} < \frac{3m}{2}  
\,.\eeq

\subsection{A general solution for equal couplings \label{sec:GeneralPlusPlus}}

In each case considered above, the solution is obtained by simple algebraic manipulations, and may as such be generalised to any $A$ and $B$. We will find the general solution in the slightly restricted case of equal couplings: $t_1=t_2=t $.
The  two saddle-point equations \eqref{eq:eig_densities_real} and \eqref{eq:eig_densities_real1} then collapse into one: 
\begin{align}
\label{eq:eig_densities_special_case}
 2\rho(\mu)    + \rho(\mu+ m) +  \rho(\mu- m) =   &      
\frac{1}{ t } 
,\end{align}
since the eigenvalue densities are equal by symmetry reasons: $\rho=\rho_\mu=\rho_\nu$.

\begin{figure} 
\centering
\includegraphics[width=0.7\textwidth]{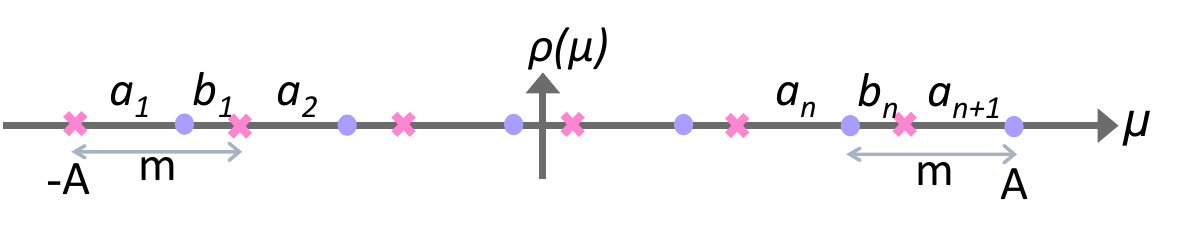}
\caption{{\small The structure of resonances in the $n$-th phase. The eigenvalue density is constant on the interval between each pair of adjacent resonances.}}
\label{fig:gen_sol_odd}
\end{figure}

Consider a solution on the interval $[-A,A]$. 
If $2A \geq m$, the endpoints will create resonances in the middle of the eigenvalue distribution at $\mp A \pm m$, and those will in turn create secondary resonances, and so the density will have discontinuities at points $\mp A \pm l m$ as long as $A-l m \geq -A$. The number of resonances thus depends on how many times the mass $m$ fits inside the length of the interval $[-A,A]$ (fig.~\ref{fig:gen_sol_odd}). This number, which we denote by $n$, characterises the different phases of the system. Within each phase, the free energy of the matrix model changes continuously with the parameters ($m$ and $t$), while it at the transition points has singularities, the exact nature of which is to be determined later.
The transition between two phases happens when the length of the interval $[-A,A]$ is an integer multiple of $m$, at which point the purple circles and pink crosses in fig.~\ref{fig:gen_sol_odd} collide. The fractional part of $2A/m$, which we denote by $\Delta $, governs the proximity to the critical point, and we thus have:
\beq
\Delta = \left\{ \frac{2A}{m} \right\} m:
\qquad 2A=m+\Delta ,
\qquad 
n=\left[\frac{2A}{m} \right]
.\eeq

Just as in the examples presented in section \ref{sec:ExamplesPlusPlus}, the eigenvalue density is constant between the resonance points. This allows us to write down an Ansatz for $\rho(\mu)$:
\begin{align}
\label{eq:rho_gen_Ansatz_PlusPlus}
\rho(\mu) =& 
\begin{cases}
a_l \hspace*{8mm} \text{for   } \mu \in [-A+m(l-1),-A+m(l-1)+ \Delta] \\
 b_l \hspace*{8mm} \text{for   } \mu \in [-A+m(l-1)+ \Delta,-A+ml] 
 \end{cases}
.\end{align}
Since for any point $\mu \in [-A+m(l-1),-A+m(l-1)+ \Delta] $, the points $\mu \pm m$ lie in the interval $ [-A+m(l-1\pm1),-A+m(l-1\pm1)+ \Delta] $,  the equations for the $a_l$ and $b_l$'s decouple, so these constants each fulfil the recursion relations:
\begin{align}
\label{eq:rec_rels}
2a_l +a_{l-1}+ a_{l+1} =& \frac{1}{ t } 
\\ \nonumber
2b_l +b_{l-1}+ b_{l+1} =& \frac{1}{t } 
,\end{align}
together with the boundary conditions that the $a_l, b_l$'s must vanish outside the support of $\rho(\mu)$, or equivalently: 
\begin{align}
a_0=b_0=0=a_{n+2}=b_{n+1}
.\end{align}
Notice that while the equations satisfied by $a_l$ and $b_l$ are the same, the boundary conditions for them differ.

The solutions to the equations in \eqref{eq:rec_rels} is  conveniently expressed in terms of a parameter
\begin{equation}
 \xi \equiv n \hspace*{2mm}(\mathop{\mathrm{mod}} 2),
\end{equation}
which gives us:
\begin{align}\label{AandB}
a_l =&  \frac{1}{4  t} \Big(1-(-1)^l\Big)  +  (-1)^l  \frac{ \xi  l }{2 t (n+2)} \\ \nonumber
b_l =&  \frac{1}{4  t} \Big(1-(-1)^l\Big)  + (-1)^l    \frac{ ( 1-\xi) l }{2 t(n+1)}
.\end{align}
Here, the constant terms are fixed by the boundary conditions at $l=0$ and the coefficients in front of the linear terms by the boundary conditions at $l=n+2$ and $l=n+1$ respectively.

There are consistency requirements on the obtained solutions: the density must be symmetric around the origin, and positive-definite or zero.
It is not immediately clear that these conditions are satisfied, but they can be readily verified by a back-of-an-envelope calculation, which we will not present here.

Using the normalisation condition of the eigenvalue density, one may find an expression for $\Delta$ in terms of the coupling parameter, the mass, and the integer $n$. The normalisation condition of $\rho(\mu)$ takes the form:
\begin{align}\label{norcond-disc}
 1&=\int_{-A}^{A}d\mu \,\rho (\mu )=
 \Delta \sum_{l=1}^{n+1} a_l+
(m- \Delta)  \sum_{l=1}^{n} b_l 
\\ \nonumber &
= 
\frac{n+2-\xi }{4t} \left[ \frac{\Delta \left(n+2+\xi \right)}{n+2} 
+\frac{\left(m-\Delta \right)\left(n+\xi \right)}{n+1}
\right] 
,\end{align}
which gives:
\beq
\label{eq:Delta}
\Delta =
 4t\,\frac{n+1+\xi }{n+2-\xi }-m(n+2\xi ),
\eeq
\begin{align}
\label{eq:int_ends}
A=2t\,\frac{n+1+\xi }{n+2-\xi }-m\xi\,.
\end{align}

\subsubsection{Phase transitions}

These expressions, \eqref{eq:Delta} and \eqref{eq:int_ends}, are not valid for any value of $n$ and $t$, since $\Delta $ by definition satisfies
\beq
0<\Delta<m
,\eeq
which originates from the fact that $\Delta$ was given by the fractional part of number of times $m$ fits inside the interval $[-A,A]$. When $\Delta $ approaches zero, the $a$-type intervals shrink to zero size (fig.~\ref{fig:gen_sol_odd}).  This happens at the critical point characterized by coupling
 \beq
t_c^n=\frac{m\left(n+2\xi \right)\left(n+2-\xi \right)}{4\left(n+1+\xi \right)}
=\frac{ m (n+\xi)(n+2-\xi )}{4(n+1)} 
.\eeq
The solution with a given $n$ thus exists for $t_c^n<t<t_c^{n+1}$. When the coupling approaches the upper critical value, the $b$-type intervals shrink, leading to the $n\rightarrow n+1$ transition. The first few critical couplings are shown in table \ref{tab:crit_coupling}. At large $n$, $t_c$ grows linearly with $n$, such that  $t_c^n\simeq mn/4$.
\begin{table}[h!]
  \begin{center}
    \begin{tabular}{| l |l l  l l l l l l l l l |}
      \hline
    $n$ & 0 &  1 & 2& 3& 4& 5& 6& 7& 8 & 9& 10\\
    \hline
    $t_c^{n}$ 
    & $0$
    & $\frac{  1}{2}  $
    & $\frac{  2}{3} $ 
    & $ 1 $ 
    & $ \frac{ 6}{ 5}  $ 
    & $ \frac{ 3}{ 2}  $ 
    & $ \frac{ 12}{ 7}  $ 
    & $ 2  $
    & $ \frac{ 20}{ 9}  $ 
    & $ \frac{ 5}{ 2}  $ 
    & $ \frac{ 30}{11}  $ \\
      \hline
     \end{tabular}
  \end{center}
  \caption{The value of the first few critical coupling parameters and the corresponding values of $n$. }
  \label{tab:crit_coupling}
\end{table}

To find the eigenvalue density for a given coupling $t$,  we first need to identify the interval $[t_c^n,t_c^{n+1}]$ within which $t$ falls. This determines $n$ and through (\ref{eq:int_ends}),\eqref{eq:rho_gen_Ansatz_PlusPlus}  and \eqref{AandB}, also the endpoints of the eigenvalue distribution as well as the density. 

At large coupling, $A$ grows asymptotically linearly with $t$: $A\simeq 2t$, but is however not a continuous function thereof. Rather, it has mild singularities at all the critical points, whose precise nature will be discussed shortly.

\begin{figure}[t]
\begin{center}
\subfigure[]{
   \includegraphics[width=0.45\textwidth] {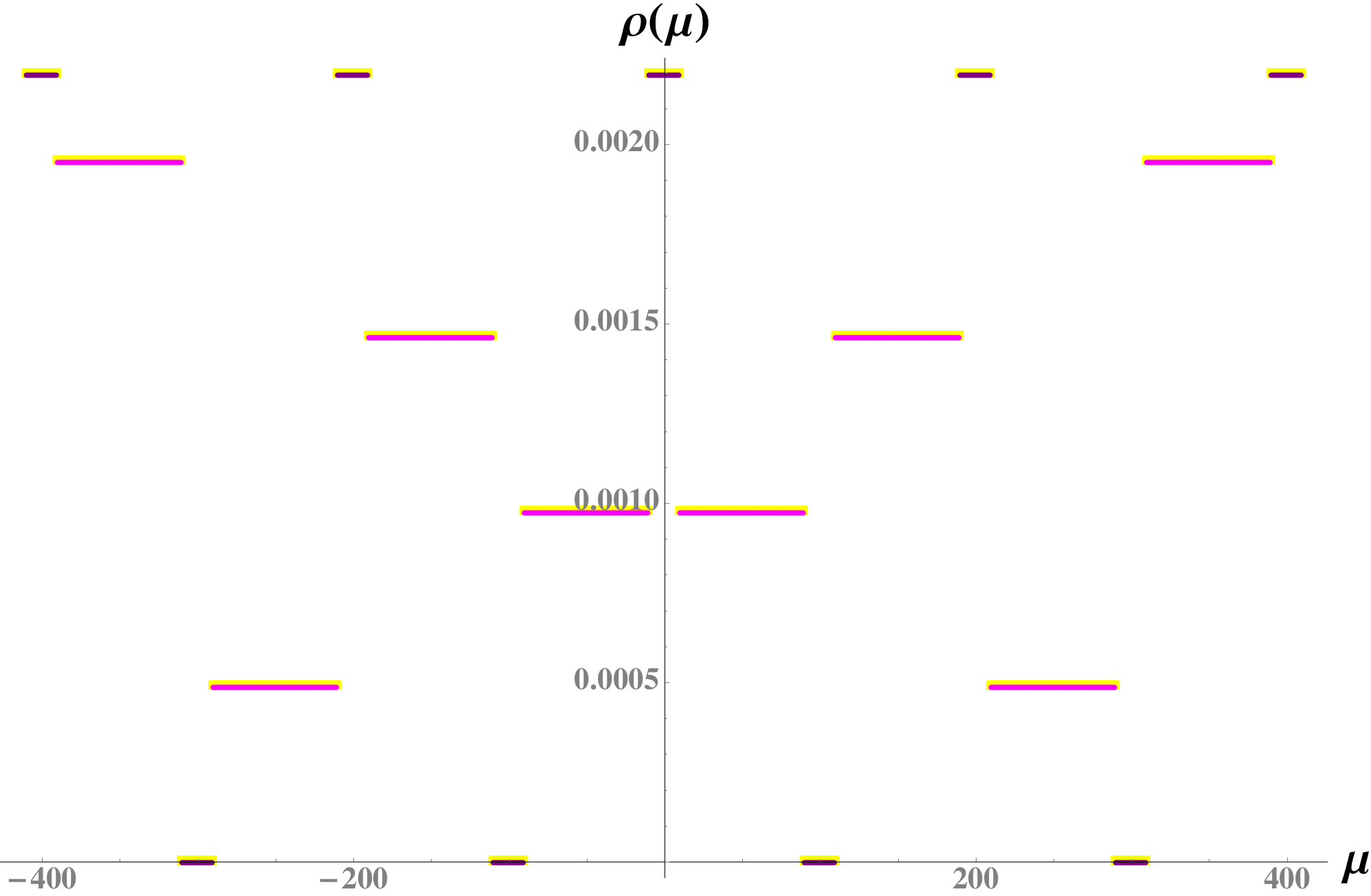}
   \label{fig:subfig1}
 }
 \subfigure[]{
   \includegraphics[width=0.45\textwidth] {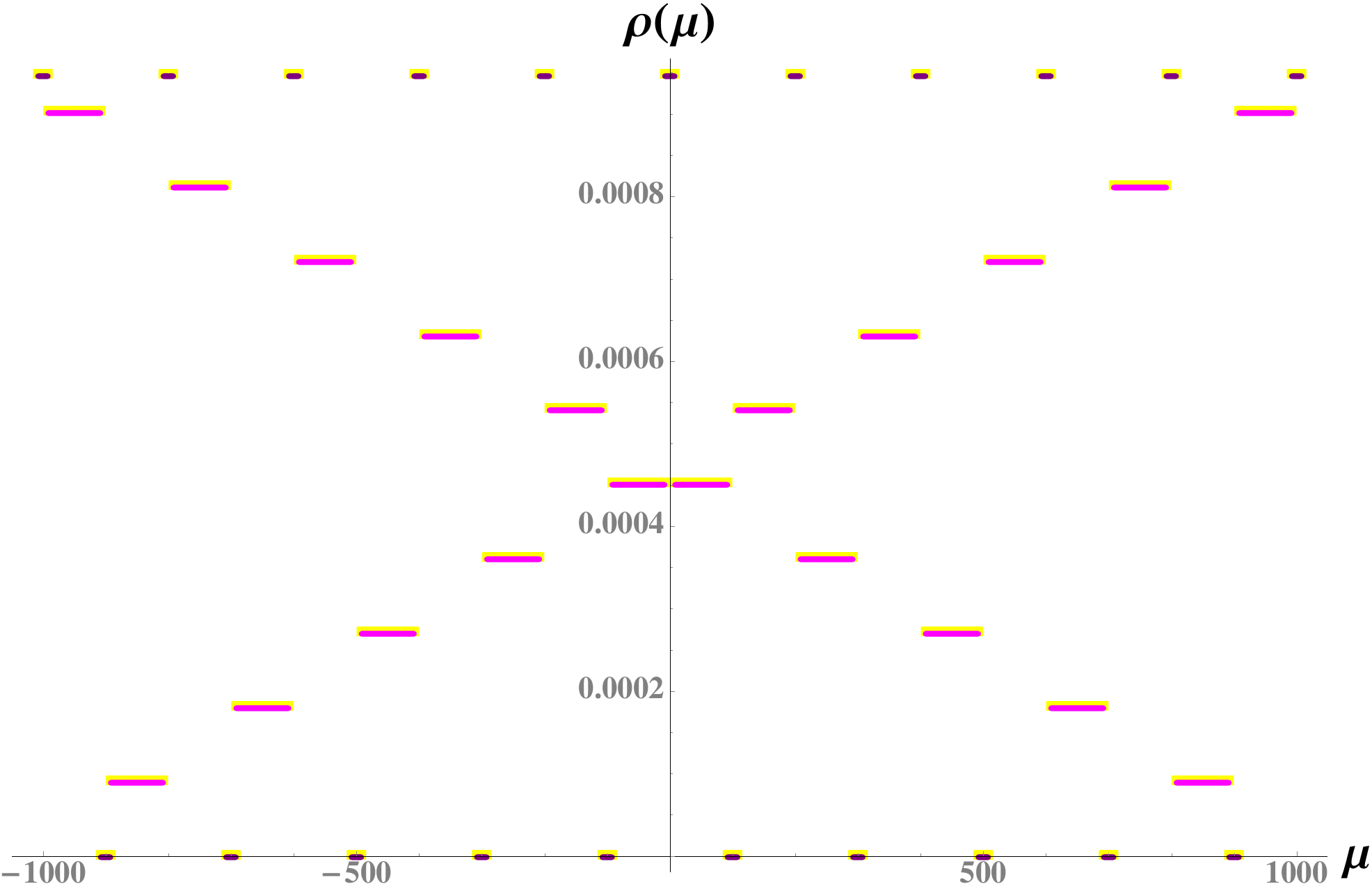}
   \label{fig:subfig2}
 }
\caption{\small  The eigenvalue density 't Hooft coupling (a) $t=41/18$, corresponding to $n=8$, and (b) $t=95/18$, corresponding to $n=20$. In  both plots $m=100$.}
\end{center}
\label{fig:n820}
\end{figure}

The first three phases in table~\ref{tab:crit_coupling} have already been discussed in
sec.~\ref{sec:ExamplesPlusPlus}. Solutions for $t=41/18$ and $t=95/18$, which correspond  to $n=8$ and $n=20$ respectively, are shown in fig.~\ref{fig:n820}.
The eigenvalue density appears to be bounded from above by $\frac{1}{2 t}$ and there are always subintervals on which the density turns to zero. In between, the constant patches align themselves along a regular, cross-like structure. 

We may also compare the analytic results in the decompactification limit with the direct numerical solutions of the saddle-point equations (\ref{eq:saddlepoints_partial_bac}). The eigenvalue density obtained numerically indeed makes plateaux which pretty well match the analytic predictions, while the corners of the steps are rounded up by finite-size effects.

\begin{figure}[t]
\begin{center}
\subfigure[]{
   \includegraphics[width=0.45\textwidth] {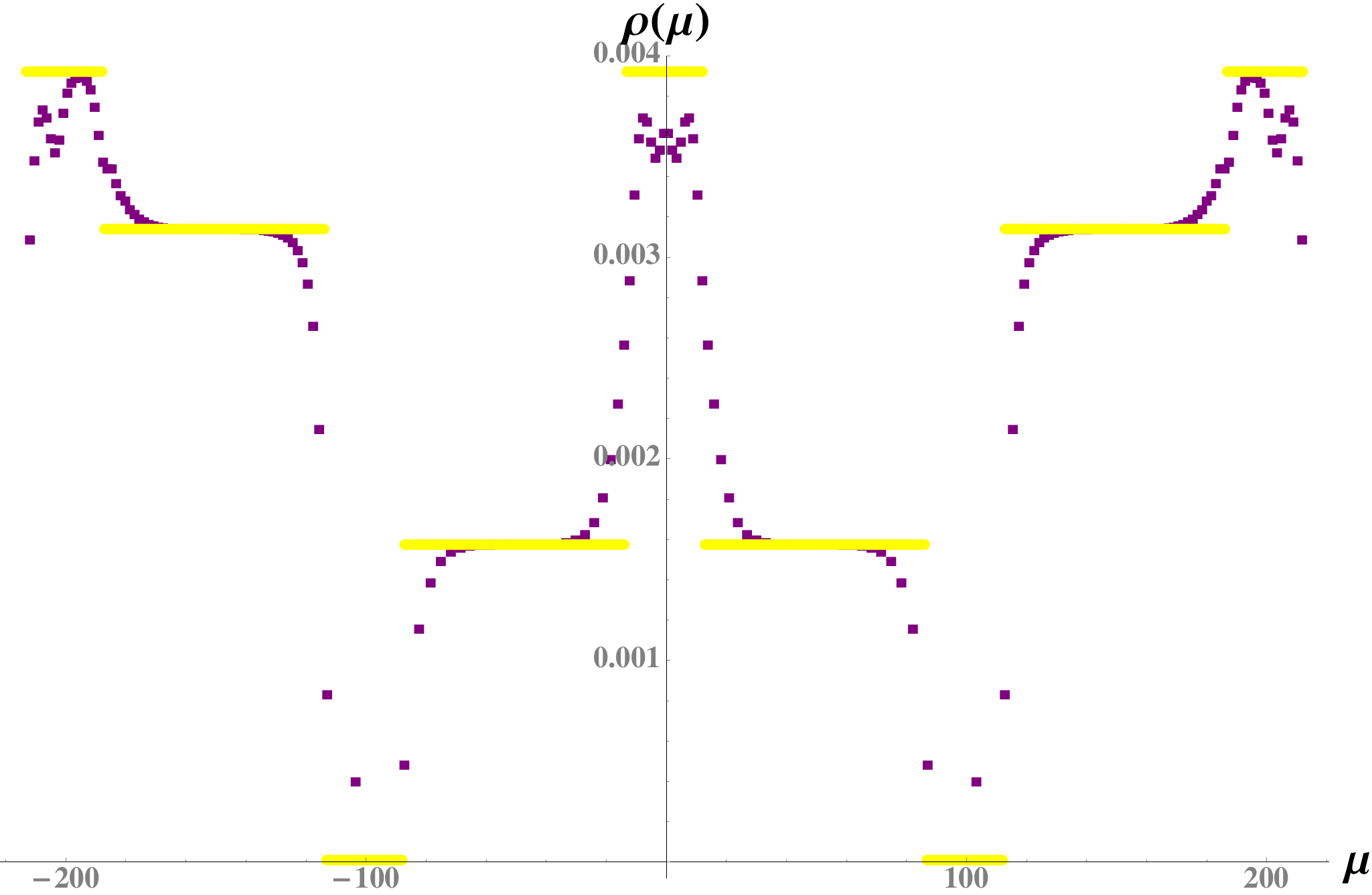}
   \label{fig:subfig1}
 }
 \subfigure[]{
   \includegraphics[width=0.45\textwidth] {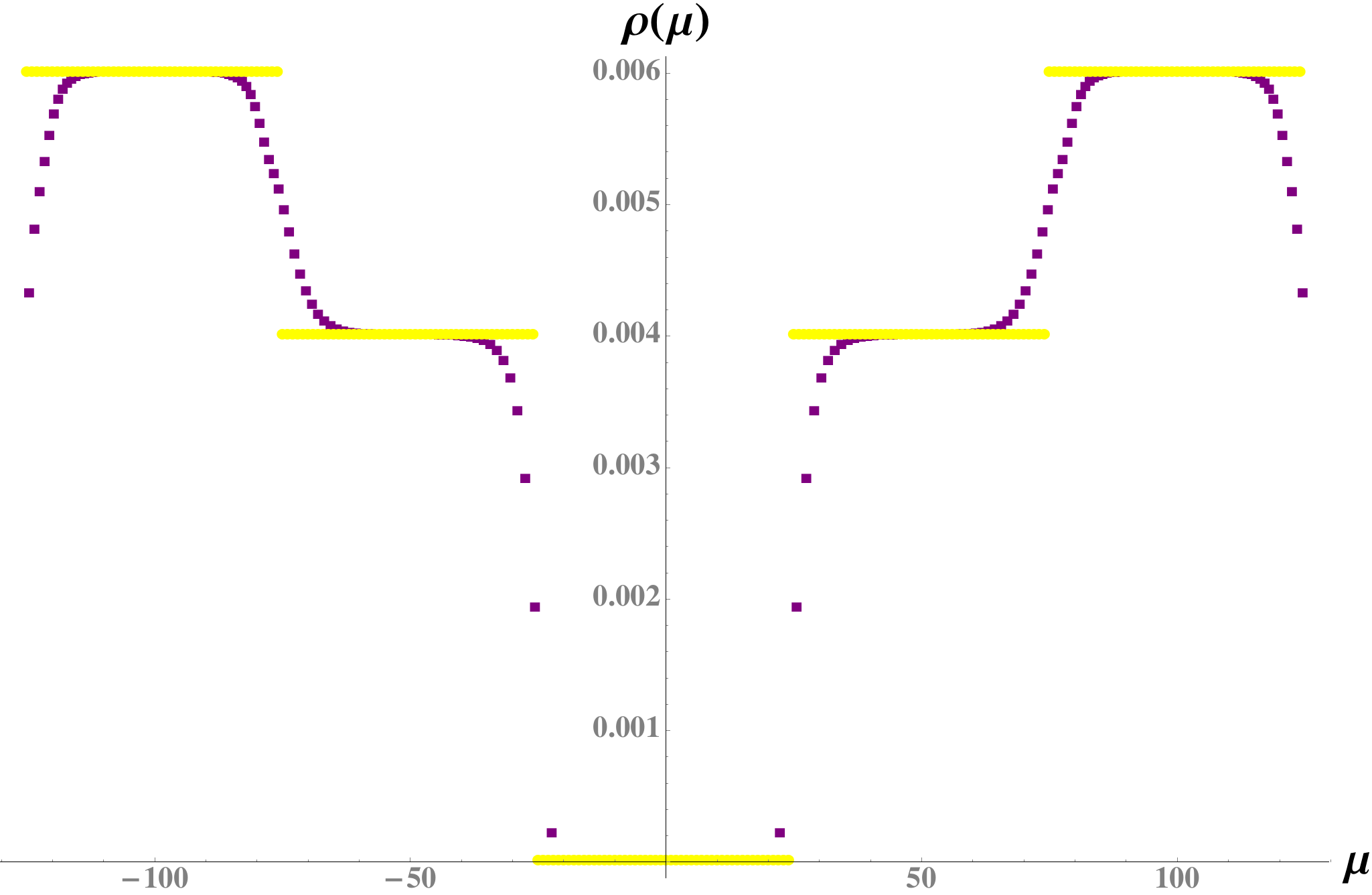}
   \label{fig:subfig2}
 }
\caption{\small  The numerically obtained eigenvalue density for $m=100$ and 't Hooft coupling (a) $t=23/18$, (corresponding to $n=4$) and (b) $t=5/6$, (corresponding to $n=2$), compared to the analytically obtained eigenvalue density. Numerical results are plotted in purple and analytical in yellow.}
\end{center}
\label{fig:comp_numerics}
\end{figure}

\subsection{Critical behavior}

The ABJM matrix model thus undergoes an infinite series of phase transitions, as a function of  the 't~Hooft coupling.
An interesting question is of what order these transitions are. To answer this, we should study how the free energy,
\begin{align}
F=-\frac{1}{R N_1N_2} \log Z
,\end{align}
changes across the critical point $t_c^n$ that separates the $(n-1)$-th and $n$-th phases. 
The free energy itself of course is continuous, but its derivatives should have discontinuities. 
It is actually easier to compute the heat capacity -- a derivative of the free energy with respect to the coupling constant,  directly. For the case where the partial 't Hooft couplings of the theory are equal, this is given by\footnote{Here $\partial _t$ should be understood as $\partial _tF=(\partial /\partial t_1+\partial /\partial t_2)F|_{t_1=t_2=t}$. Notice that for unequal couplings, $\partial F/\partial t_{1,2}$ cannot be expressed as a local integral of the density.}:
 \begin{align}
\partial_t F=-\frac{\langle \mu^2 \rangle}{ t^2 }  = -\frac{1}{ t^2 } \int_{-A}^{A}  d \mu\,\rho(\mu) \mu^2
 .\end{align}

Using the explicit form of the eigenvalue density, the heat capacity can be written as
 \begin{align}
-3t^2\partial_t F & =
\sum_{l=1}^{n+1}  a_l \left\{
\left[ -A+m(l-1)+ \Delta\right]^3 - \left[-A+m(l-1)\right]^3
\right\}
\\ \nonumber &
\hspace*{2cm}
+
\sum_{l=1}^{n} b_l 
\left\{ (-A+m l)^3-\left[-A+m(l-1)+ \Delta\right]^3 
\right\}
 ,\end{align}
allowing us to compute the derivatives of the free energy at the critical coupling. Denoting the free energy in the $n^{\text{th}}$ phase by  $F_n$,  
after some algebra
we obtain:
\begin{align}
\partial_t (F_{n-1}-F_{n})|_{t_c^n} = \partial^2_t (F_{n-1}-F_{n})|_{t_c^n} =& 0 
,\end{align}
whereas the third derivative of the free energy experiences a finite jump:
\begin{align}
\partial^3_t (F_{n-1}-F_{n})|_{t_c^n}=
\begin{cases}
&
-\frac{512 (1+n)^5}{n^4 (2+n)^4m^2} \hspace*{2.4cm } \text{even } n \\
&
\frac{512 m}{(1+n)^3m^2} \hspace*{3cm } \text{odd } n
\end{cases}
.\end{align}
If we do a similar comparison for the endpoint of the interval, with the help of (\ref{eq:int_ends}), we find that $A$ itself is continuous, while its first derivative experiences a jump at the critical point.

Thus we conclude that there is an infinite number of  quantum phase transitions in the mass-deformed ABJM model, all of  third order. Furthermore, these transitions become weaker and weaker with increasing coupling, which may be easily seen by noting that the discontinuity in the third derivative of the free energy scales as $1/n^3$ for large $n$.
 
\section{ Another analytic continuation \label{sec:ContLevel}} 

It is interesting to compare the behavior that we have found for the ABJM matrix model continued in the rank of the gauge group with the model (\ref{eq:matrix_model_scaled_acont}), obtained by analytic continuation in the Chern-Simons level. This will allow us to study the quantum weak/strong phase transitions which were found in section \ref{sec:exact-s} from a different perspective. The pattern that we will find in the decompactification limit turns out to be strikingly similar to the behavior of the $\mathcal{N}=2^*$ theory in four dimensions \cite{Russo:2013qaa,Russo:2013kea,Russo:2013sba}: the eigenvalue density at strong coupling, as we shall see, has an enveloping limit shape, with a fine irregular structure on top. As in the previous section, we will be able to resolve this fine structure analytically at any value of the coupling and thus map the entire phase diagram of the model.

Again, in the decompactification limit we need to scale the couplings, $\alpha _{1,2}$, with $R$,
\begin{equation}
 \alpha _{1,2}=Rg_{1,2},
\end{equation}
such that $g_{1,2}$ remain finite when $R\rightarrow \infty $. The saddle-point equations, (\ref{eq:saddlepoints_partial_bac_acont}) after the same steps that led to \eqref{eq:eig_densities_real}, turn to finite-difference equations:
\begin{align}
\label{fdiff-eq}
 \rho_\mu(\mu)    =   &      
\frac{1}{2 g_1  }
+  \frac{1}{2  }      \Big( \rho_\nu(\mu+ m)  +  \rho_\nu(\mu- m) \Big) \\
 \rho_\nu(\nu)  =   &       
\frac{1}{ 2 g_2 }
-\frac{1}{2   }     \Big(  \rho_\mu(\nu+m) +  \rho_\mu(\nu-m)   \Big) 
.\end{align}
In the case of equal couplings, $g_1=g_2=g$, those two equations reduce to one:
\begin{align}
\label{eq:eig_densities_special_case2}
 2\rho(\mu)    - \rho(\mu+ m) - \rho(\mu- m) =   &      
\frac{1}{ g }\, 
,\end{align}
which differs from (\ref{eq:eig_densities_special_case}) by two signs.

At strong coupling,  $g \ll m$, the difference operator in (\ref{eq:eig_densities_special_case2}) becomes differential. Then, approximately, $-m^2\rho ''=1/g$, which is solved by
\begin{equation}\label{parab}
 \rho_\infty  (\mu)=\frac{1}{2 g m^2}\left(A_\infty ^2-\mu^2\right)
,\end{equation}
supported on the interval $[-A_\infty,A_\infty]$, where the interval endpoints are determined by the normalisation condition:
\begin{equation}\label{rho-inf}
A_\infty  =\left(\frac{3 g m^2}{2}\right)^{\frac{1}{3}}.
\end{equation}
The infinite-coupling solution is a smooth envelope of a rigged, irregular structure on small scales \cite{Russo:2013qaa,Russo:2013kea}. Below, we will find an analytic solution that describes this fine structure of the density.

The solution largely follows the analysis in sec.~\ref{sec:GeneralPlusPlus}.
Parametrising the density as in (\ref{eq:rho_gen_Ansatz_PlusPlus}), we find two sets of identical recursion relations for the constants $a_l$ and $b_l$:
\begin{equation}
 2a_l-a_{l+1}-a_{l-1}=\frac{1}{g}\,,\qquad  2b_l-b_{l+1}-b_{l-1}=\frac{1}{g}\,,
\end{equation}
supplemented by the boundary conditions (\ref{eq:rec_rels}).
The solution to these equations is
\begin{eqnarray}
 a_l&=&\frac{1}{2 g}\left[\left(n+2\right)l^2-l^2\right], \qquad l=1\ldots n+1
 \\
 b_l&=&\frac{1}{2 g}\left[\left(n+1\right)l^2-l^2\right], \qquad l=1\ldots n.
\end{eqnarray}
The normalization condition, in the form (\ref{norcond-disc}),
fixes $\Delta $ and hence $A $:
\begin{equation}\label{Delta-mu}
 \Delta =\frac{4g }{\left(n+1\right)\left(n+2\right)}-\frac{nm}{3}\,,\qquad 
A=\frac{nm}{3}+\frac{2g}{\left(n+1\right)\left(n+2\right)}\,.
\end{equation}

Just as in the previous case, the two types of resonances, (those associated with the right end of the eigenvalue distribution and those associated with the left end), collide when $\Delta \rightarrow 0$ or $\Delta \rightarrow 1$. The $a$ or $b$ cuts then shrink, and the system undergoes a transition to a new phase with different $n$. From (\ref{Delta-mu}), we find that this transition happens at
\begin{equation}
g_c^{n}=\frac{mn\left(n+1\right)\left(n+2\right)}{12}\,.
\end{equation}
At $g_c^1={m}/{2}$, the first two resonances appear near the endpoints of the distribution, signalling a transition to the $n=1$ phase where the density has three patches. As the coupling parameter  increases further, the resonances move into the interior of the eigenvalue distribution, and at $g_c^{2}=2m$, these resonances collide, while two new resonances are nucleated near the endpoints, and so on.

\begin{figure}[t]
\begin{center}
 \centerline{\includegraphics[width=9cm]{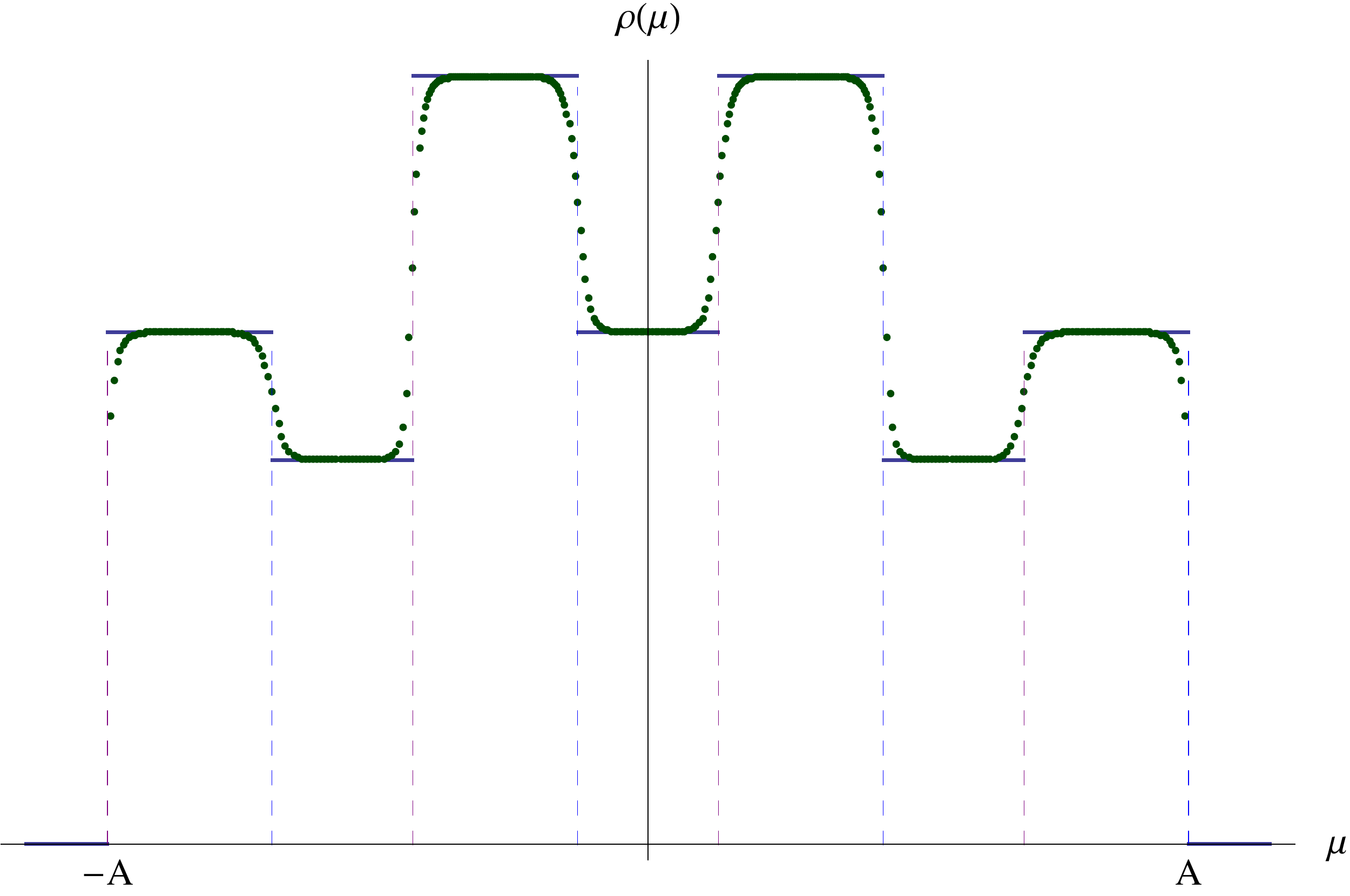}}
\caption{\label{ex-vs-infinite}\small  Comparison of the infinite-volume density (blue lines) with the solution of the finite-volume saddle-point equations (\ref{eq:saddlepoints_partial_bac_acont}) (green dots) for sufficiently big $m$. }
\end{center}
\end{figure}

\begin{figure}[t]
\begin{center}
 \centerline{\includegraphics[width=9cm]{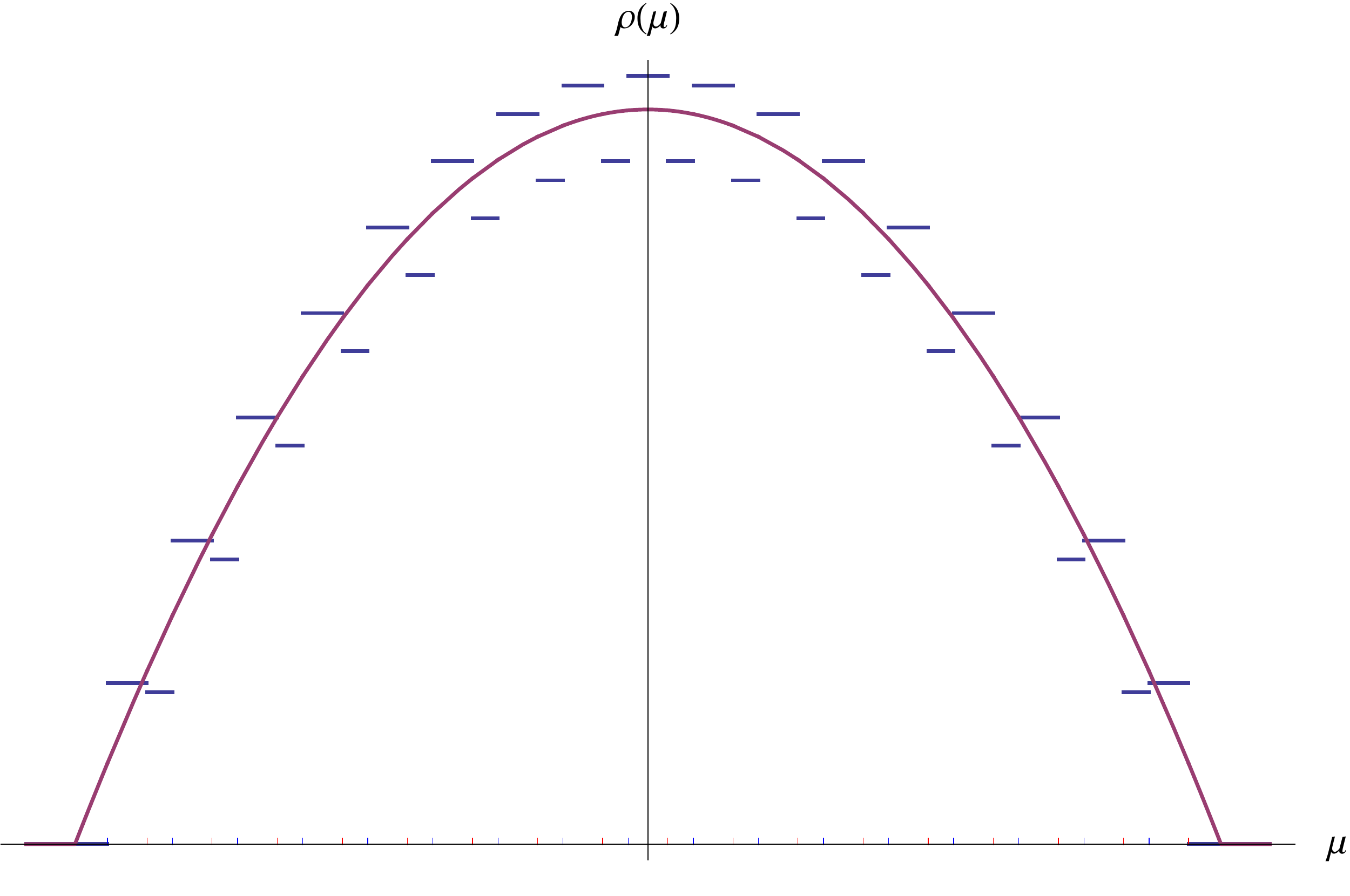}}
\caption{\label{comparison-to-rho-infinity}\small  The exact solution at sufficiently large coupling (blue lines) vs. the limiting shape (\ref{parab}).}
\end{center}
\end{figure}

The solution can be compared with numerical results for the finite-volume model, fig.~\ref{ex-vs-infinite}. It is also possible to see that at large coupling, ($g\rightarrow \infty $), the exact eigenvalue density approaches the limiting parabolic shape (\ref{parab}), fig.~\ref{comparison-to-rho-infinity}.

One can, once more, compute the heat capacity, using the formula:
\begin{eqnarray}
 \left\langle \mu ^2\right\rangle
 &=&
 \frac{4g^2}{3\left(n+1\right)^2\left(n+2\right)^2}
 +\frac{n\left(n+3\right)m^2}{18}
\nonumber \\
&&
-\frac{n\left(n+1\right)\left(n+2\right)\left(n+3\right)\left(2n+3\right)m^3}{1620g}\,.
\end{eqnarray}
It is then straight-forward to show that the third derivative of the free energy exhibits a discontinuity at $g=g_c^n$, whereas all lower-order derivatives are continuous at all values of the coupling. Consequently all of the transitions are of third order with zero critical indices, as in the previous case.

\section{Discussion}\label{sec:discussion}

We have herein studied two analytically continued versions of the mass-deformed ABJM theory at large $N$, whose partition function on the sphere can be computed with the help of localisation.  Due to the simplicity of the saddle-point equations, the phase structure of the model can be completely mapped out at any coupling.
In both analytic continuations of the model, an infinite series of phase transitions, located at some critical values of the coupling, is found. Furthermore, these phase transitions accumulate as the coupling becomes infinite, which  raises the question of what can be a holographic dual of this fractal structure.

As in many papers on localisation in three dimensions, we solved the model with the couplings (either the rank of the gauge group of the Chern-Simons level) analytically continued into the complex plane. In contrast to conformal theories, where analytic continuation back to the physical couplings is straightforward, here such an analytic continuation actually poses a problem. This, in part, is due to the fact that the hyperbolic functions in the saddle-point equations \eqref{eq:saddlepoints_partial_bac} and \eqref{eq:saddlepoints_partial_bac_acont} develop poles as the arguments obtain imaginary parts. 

Problems with analytic continuation are readily visible in the saddle-point equations (\ref{eq:saddlepoints_partial_bac_acont}). Suppose that we are interested in the regime when the mass is big, and all eigenvalues are big. We know that for real $\alpha _1$, $\alpha _2$, the eigenvalue  density forms a number of steps of different hight along the real axis. We now want to rotate $\alpha _{1,2}$ into the complex plane. The eigenvalues will also become complex, but the hyperbolic $\tanh$ and $\coth$ are to first approximations real-valued step-functions irrespectively of whether their argument is real or complex. The right hand side of \eqref{eq:saddlepoints_partial_bac_acont}  is thus approximately real-valued, and so should be the left-hand-side, which means that all the $\mu $ (respectively, $\nu $) eigenvalues have the same complex phase. But this is clearly inconsistent with the density taking different values on different patches of the eigenvalue distribution. Our preliminary numerical analysis shows that the solution becomes very unstable for complex couplings, and does not form one-dimensional lines on the complex plane, as in the usual matrix model \cite{David:1990sk}, but rather resembles a random scatter plot. This is true for both types of analytic continuation we have considered. The way to transfer the results obtained herein back to the original theory in a satisfactory fashion remains a mystery and would be an interesting area for future work.

It would furthermore be interesting to understand if the models that we considered here have hologrpahic duals. The holographic dual of the mass-deformed ABJM theory is not known. In principle, it can be constructed by switching on a constant source for the supergravity field dual to the mass operator, and following the flow triggered by this source. A related construction was studied in \cite{Cheon:2011gv}.


\section*{Acknowledgement}

We would like to thank J.~K\"all\'{e}n for very useful discussions and J.~Russo for comments on the manuscript.
The work of K.Z. was supported by the Marie
Curie network GATIS of the European Union's FP7 Programme under REA Grant
Agreement No 317089, by the ERC advanced grant No 341222
and by the Swedish Research Council (VR) grant
2013-4329. 

\bibliographystyle{nb}

\begin{thebibliography}{10}
\ifx\href\asklfhas\newcommand{\href}[2]{#2}\fi
\raggedright
\small
\parskip 0pt

\bibitem{Witten:1988ze}
E.~Witten,
\textit{``{Topological Quantum Field Theory}''},
\textsf{Commun.Math.Phys.~117,~353~(1988)}.
%
\bibitem{Pestun:2007rz}
V.~Pestun,
\textit{``{Localization of gauge theory on a four-sphere and supersymmetric
  Wilson loops}''},
\textsf{Commun.Math.Phys.~313,~71~(2012)},
\href{http://arXiv.org/abs/0712.2824}{\texttt{0712.2824}}.
%
\bibitem{Brezin:1977sv}
E.~Brezin, C.~Itzykson, G.~Parisi and J.~B.~Zuber,
\textit{``{Planar Diagrams}''},
\textsf{Commun.~Math.~Phys.~59,~35~(1978)}.
%
\bibitem{Aharony:2008ug}
O.~Aharony, O.~Bergman, D.~L.~Jafferis and J.~Maldacena,
\textit{``{N=6 superconformal Chern-Simons-matter theories, M2-branes and their
  gravity duals}''},
\textsf{JHEP~0810,~091~(2008)},
\href{http://arXiv.org/abs/0806.1218}{\texttt{0806.1218}}.
%
\bibitem{Kapustin:2009kz}
A.~Kapustin, B.~Willett and I.~Yaakov,
\textit{``{Exact Results for Wilson Loops in Superconformal Chern-Simons
  Theories with Matter}''},
\textsf{JHEP~1003,~089~(2010)},
\href{http://arXiv.org/abs/0909.4559}{\texttt{0909.4559}}.
%
\bibitem{Kapustin:2010xq}
A.~Kapustin, B.~Willett and I.~Yaakov,
\textit{``{Nonperturbative Tests of Three-Dimensional Dualities}''},
\textsf{JHEP~1010,~013~(2010)},
\href{http://arXiv.org/abs/1003.5694}{\texttt{1003.5694}}.
%
\bibitem{Drukker:2010nc}
N.~Drukker, M.~Marino and P.~Putrov,
\textit{``{From weak to strong coupling in ABJM theory}''},
\textsf{Commun.Math.Phys.~306,~511~(2011)},
\href{http://arXiv.org/abs/1007.3837}{\texttt{1007.3837}}.
%
\bibitem{Herzog:2010hf}
C.~P.~Herzog, I.~R.~Klebanov, S.~S.~Pufu and T.~Tesileanu,
\textit{``{Multi-Matrix Models and Tri-Sasaki Einstein Spaces}''},
\textsf{Phys.Rev.~D83,~046001~(2011)},
\href{http://arXiv.org/abs/1011.5487}{\texttt{1011.5487}}.
%
\bibitem{Martelli:2011qj}
D.~Martelli and J.~Sparks,
\textit{``{The large N limit of quiver matrix models and Sasaki-Einstein
  manifolds}''},
\textsf{Phys.Rev.~D84,~046008~(2011)},
\href{http://arXiv.org/abs/1102.5289}{\texttt{1102.5289}}.
%
\bibitem{Cheon:2011vi}
S.~Cheon, H.~Kim and N.~Kim,
\textit{``{Calculating the partition function of N=2 Gauge theories on $S^3$
  and AdS/CFT correspondence}''},
\textsf{JHEP~1105,~134~(2011)},
\href{http://arXiv.org/abs/1102.5565}{\texttt{1102.5565}}.
%
\bibitem{Jafferis:2011zi}
D.~L.~Jafferis, I.~R.~Klebanov, S.~S.~Pufu and B.~R.~Safdi,
\textit{``{Towards the F-Theorem: N=2 Field Theories on the Three-Sphere}''},
\textsf{JHEP~1106,~102~(2011)},
\href{http://arXiv.org/abs/1103.1181}{\texttt{1103.1181}}.
%
\bibitem{Drukker:2011zy}
N.~Drukker, M.~Marino and P.~Putrov,
\textit{``{Nonperturbative aspects of ABJM theory}''},
\textsf{JHEP~1111,~141~(2011)},
\href{http://arXiv.org/abs/1103.4844}{\texttt{1103.4844}}.
%
\bibitem{Fuji:2011km}
H.~Fuji, S.~Hirano and S.~Moriyama,
\textit{``{Summing Up All Genus Free Energy of ABJM Matrix Model}''},
\textsf{JHEP~1108,~001~(2011)},
\href{http://arXiv.org/abs/1106.4631}{\texttt{1106.4631}}.
%
\bibitem{Gulotta:2011aa}
D.~R.~Gulotta, C.~P.~Herzog and S.~S.~Pufu,
\textit{``{Operator Counting and Eigenvalue Distributions for 3D Supersymmetric
  Gauge Theories}''},
\textsf{JHEP~1111,~149~(2011)},
\href{http://arXiv.org/abs/1106.5484}{\texttt{1106.5484}}.
%
\bibitem{Alday:2012au}
L.~F.~Alday, M.~Fluder and J.~Sparks,
\textit{``{The Large N limit of M2-branes on Lens spaces}''},
\textsf{JHEP~1210,~057~(2012)},
\href{http://arXiv.org/abs/1204.1280}{\texttt{1204.1280}}.
%
\bibitem{Assel:2012cp}
B.~Assel, J.~Estes and M.~Yamazaki,
\textit{``{Large N Free Energy of 3d N=4 SCFTs and $AdS_4/CFT_3$}''},
\textsf{JHEP~1209,~074~(2012)},
\href{http://arXiv.org/abs/1206.2920}{\texttt{1206.2920}}.
%
\bibitem{Bhattacharyya:2012ye}
S.~Bhattacharyya, A.~Grassi, M.~Marino and A.~Sen,
\textit{``{A One-Loop Test of Quantum Supergravity}''},
\textsf{Class.Quant.Grav.~31,~015012~(2014)},
\href{http://arXiv.org/abs/1210.6057}{\texttt{1210.6057}}.
%
\bibitem{Freedman:2013oja}
D.~Z.~Freedman and S.~S.~Pufu,
\textit{``{The holography of $F$-maximization}''},
\textsf{JHEP~1403,~135~(2014)},
\href{http://arXiv.org/abs/1302.7310}{\texttt{1302.7310}}.
%
\bibitem{Farquet:2013cwa}
D.~Farquet and J.~Sparks,
\textit{``{Wilson loops and the geometry of matrix models in
  AdS$_4$/CFT$_3$}''},
\textsf{JHEP~1401,~083~(2014)},
\href{http://arXiv.org/abs/1304.0784}{\texttt{1304.0784}}.
%
\bibitem{Hatsuda:2013oxa}
Y.~Hatsuda, M.~Marino, S.~Moriyama and K.~Okuyama,
\textit{``{Non-perturbative effects and the refined topological string}''},
\href{http://arXiv.org/abs/1306.1734}{\texttt{1306.1734}}.
%
\bibitem{Marmiroli:2013nza}
D.~Marmiroli,
\textit{``{Notes on BPS Wilson Loops and the Cusp Anomalous Dimension in ABJM
  theory}''},
\href{http://arXiv.org/abs/1312.2972}{\texttt{1312.2972}}.
%
\bibitem{Lewkowycz:2013laa}
A.~Lewkowycz and J.~Maldacena,
\textit{``{Exact results for the entanglement entropy and the energy radiated
  by a quark}''},
\textsf{JHEP~1405,~025~(2014)},
\href{http://arXiv.org/abs/1312.5682}{\texttt{1312.5682}}.
%
\bibitem{Bianchi:2014laa}
M.~S.~Bianchi, L.~Griguolo, M.~Leoni, S.~Penati and D.~Seminara,
\textit{``{BPS Wilson loops and Bremsstrahlung function in ABJ(M): a two loop
  analysis}''},
\href{http://arXiv.org/abs/1402.4128}{\texttt{1402.4128}}.
%
\bibitem{Gromov:2014eha}
N.~Gromov and G.~Sizov,
\textit{``{Exact Slope and Interpolating Functions in ABJM Theory}''},
\href{http://arXiv.org/abs/1403.1894}{\texttt{1403.1894}}.
%
\bibitem{Farquet:2014kma}
D.~Farquet, J.~Lorenzen, D.~Martelli and J.~Sparks,
\textit{``{Gravity duals of supersymmetric gauge theories on
  three-manifolds}''},
\href{http://arXiv.org/abs/1404.0268}{\texttt{1404.0268}}.
%
\bibitem{Honda:2014npa}
M.~Honda and K.~Okuyama,
\textit{``{Exact results on ABJ theory and the refined topological string}''},
\href{http://arXiv.org/abs/1405.3653}{\texttt{1405.3653}}.
%
\bibitem{Dabholkar:2014wpa}
A.~Dabholkar, N.~Drukker and J.~Gomes,
\textit{``{Localization in Supergravity and Quantum $AdS_4/CFT_3$
  Holography}''},
\href{http://arXiv.org/abs/1406.0505}{\texttt{1406.0505}}.
%
\bibitem{Barranco:2014tla}
A.~Barranco and J.~G.~Russo,
\textit{``{Large N phase transitions in supersymmetric Chern-Simons theory with
  massive matter}''},
\textsf{JHEP~1403,~012~(2014)},
\href{http://arXiv.org/abs/1401.3672}{\texttt{1401.3672}}.
%
\bibitem{Russo:2013kea}
J.~Russo and K.~Zarembo,
\textit{``{Massive N=2 Gauge Theories at Large N}''},
\textsf{JHEP~1311,~130~(2013)},
\href{http://arXiv.org/abs/1309.1004}{\texttt{1309.1004}}.
%
\bibitem{Russo:2013qaa}
J.~G.~Russo and K.~Zarembo,
\textit{``{Evidence for Large-N Phase Transitions in N=2* Theory}''},
\textsf{JHEP~1304,~065~(2013)},
\href{http://arXiv.org/abs/1302.6968}{\texttt{1302.6968}}.
%
\bibitem{Jafferis:2010un}
D.~L.~Jafferis,
\textit{``{The Exact Superconformal R-Symmetry Extremizes Z}''},
\textsf{JHEP~1205,~159~(2012)},
\href{http://arXiv.org/abs/1012.3210}{\texttt{1012.3210}}.
%
\bibitem{Hama:2010av}
N.~Hama, K.~Hosomichi and S.~Lee,
\textit{``{Notes on SUSY Gauge Theories on Three-Sphere}''},
\textsf{JHEP~1103,~127~(2011)},
\href{http://arXiv.org/abs/1012.3512}{\texttt{1012.3512}}.
%
\bibitem{Marino:2011nm}
M.~Marino,
\textit{``{Lectures on localization and matrix models in supersymmetric
  Chern-Simons-matter theories}''},
\textsf{J.Phys.A~A44,~463001~(2011)},
\href{http://arXiv.org/abs/1104.0783}{\texttt{1104.0783}}.
%
\bibitem{Marino:2002fk}
M.~Marino,
\textit{``{Chern-Simons theory, matrix integrals, and perturbative three
  manifold invariants}''},
\textsf{Commun.Math.Phys.~253,~25~(2004)},
\href{http://arXiv.org/abs/hep-th/0207096}{\texttt{hep-th/0207096}}.
%
\bibitem{Aganagic:2002wv}
M.~Aganagic, A.~Klemm, M.~Marino and C.~Vafa,
\textit{``{Matrix model as a mirror of Chern-Simons theory}''},
\textsf{JHEP~0402,~010~(2004)},
\href{http://arXiv.org/abs/hep-th/0211098}{\texttt{hep-th/0211098}}.
%
\bibitem{Halmagyi:2003ze}
N.~Halmagyi and V.~Yasnov,
\textit{``{The Spectral curve of the lens space matrix model}''},
\textsf{JHEP~0911,~104~(2009)},
\href{http://arXiv.org/abs/hep-th/0311117}{\texttt{hep-th/0311117}}.
%
\bibitem{Russo:2013sba}
J.~Russo and K.~Zarembo,
\textit{``{Localization at Large N}''},
\href{http://arXiv.org/abs/1312.1214}{\texttt{1312.1214}}.
%
\bibitem{Suyama:2009pd}
T.~Suyama,
\textit{``{On Large N Solution of ABJM Theory}''},
\textsf{Nucl.Phys.~B834,~50~(2010)},
\href{http://arXiv.org/abs/0912.1084}{\texttt{0912.1084}}.
%
\bibitem{Suyama:2011yz}
T.~Suyama,
\textit{``{Eigenvalue Distributions in Matrix Models for Chern-Simons-matter
  Theories}''},
\textsf{Nucl.Phys.~B856,~497~(2012)},
\href{http://arXiv.org/abs/1106.3147}{\texttt{1106.3147}}.
%
\bibitem{Giasemidis:2013oea}
G.~Giasemidis, R.~J.~Szabo and M.~Tierz,
\textit{``{Supersymmetric gauge theories, Coulomb gases and Chern-Simons matrix
  models}''},
\textsf{Phys.Rev.~D89,~025016~(2014)},
\href{http://arXiv.org/abs/1310.3122}{\texttt{1310.3122}}.
%
\bibitem{David:1990sk}
F.~David,
\textit{``{Phases of the large N matrix model and nonperturbative effects in
  2-d gravity}''},
\textsf{Nucl.Phys.~B348,~507~(1991)}.
%
\bibitem{Cheon:2011gv}
S.~Cheon, H.-C.~Kim and S.~Kim,
\textit{``{Holography of mass-deformed M2-branes}''},
\href{http://arXiv.org/abs/1101.1101}{\texttt{1101.1101}}.
%
\end{thebibliography}

\end{document}